\documentclass[submission, Proceedings]{SciPost}

\hypersetup{
    colorlinks,
    linkcolor={red!50!black},
    citecolor={blue!50!black},
    urlcolor={blue!80!black}
}

\usepackage[bitstream-charter]{mathdesign}
\usepackage[capitalise]{cleveref}
\usepackage{enumitem} 
\usepackage{wrapfig}
\DeclareSymbolFont{usualmathcal}{OMS}{cmsy}{m}{n}
\DeclareSymbolFontAlphabet{\mathcal}{usualmathcal}
\let\OLDthebibliography\thebibliography
\renewcommand\thebibliography[1]{
  \OLDthebibliography{#1}
  \setlength{\parskip}{0pt}
  \setlength{\itemsep}{1pt plus 0.3ex}
}

\begin{document}

\begin{center}{\Large\textbf{
Diffractive dissociation in future electron-ion colliders
}}\end{center}

% TODO: write the author list here. Use initials + surname format.
% Separate subsequent authors by a comma, omit comma at the end of the list.
% Mark the corresponding author with a superscript *.
\begin{center}
Anh Dung Le\textsuperscript{$\star$},
\end{center}

\begin{center}
CPHT, CNRS, {\'E}cole polytechnique, IP Paris, F-91128 Palaiseau, France
\\
* dung.le@polytechnique.edu
\end{center}

\begin{center}
\today
\end{center}

% For convenience during refereeing (optional),
% you can turn on line numbers by uncommenting the next line:
%\linenumbers

\definecolor{palegray}{gray}{0.95}
\begin{center}
\colorbox{palegray}{
  \begin{tabular}{rr}
  \begin{minipage}{0.1\textwidth}
    \includegraphics[width=22mm]{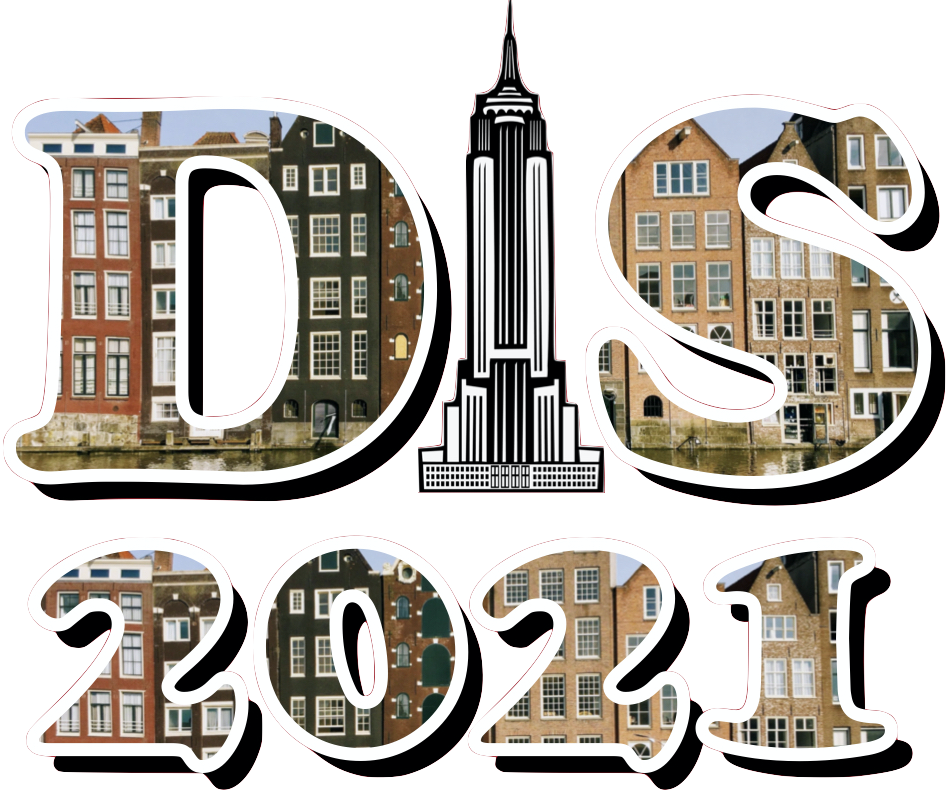}
  \end{minipage}
  &
  \begin{minipage}{0.75\textwidth}
    \begin{center}
    {\it Proceedings for the XXVIII International Workshop\\ on Deep-Inelastic Scattering and
Related Subjects,}\\
    {\it Stony Brook University, New York, USA, 12-16 April 2021} \\
    \doi{10.21468/SciPostPhysProc.?}\\
    \end{center}
  \end{minipage}
\end{tabular}
}
\end{center}

\section*{Abstract}
{\bf
% TODO: write your abstract here.
We study diffractive scattering cross sections, focusing on the rapidity gap distribution in realistic kinematics at future electron-ion colliders. Our study consists in numerical solutions of the QCD evolution equations in both fixed and running coupling frameworks. The fixed and the running coupling equations are shown to lead to different shapes for the rapidity gap distribution. The obtained distribution when the coupling is fixed exhibits a shape characteristic of a recently developed model for diffractive dissociation, which indicates the relevance of the study of that diffractive observable for the partonic-level understanding of diffraction. 
}

%\vspace{10pt}
%\noindent\rule{\textwidth}{1pt}
%\tableofcontents\thispagestyle{fancy}
%\noindent\rule{\textwidth}{1pt}
%\vspace{10pt}

\section{Introduction}
\label{sec:intro}
% TODO: write your article here.
Single diffractive dissociation in electron-hadron collisions is defined to be the scattering process in which the virtual photon mediating the interaction fluctuates into a set of partons part of which goes in the final state while the hadron remains intact, leaving a large rapidity gap between the latter and the slowest produced particle. The rapidity gap characterization of diffractive events is manifested in collision experiments by the existence of a large angular sector in the detector with no measured particle. Such events were first observed in electron-proton collisions at DESY HERA~\cite{h1.1995,zeus.1995}, and their detailed analysis is one of goals of the construction of future electron-ion colliders~\cite{LHeC.2012,EIC.2016,AbdulKhalek:2021gbh}.

Diffractive dissociation in deep-inelastic scattering (DIS) is a process of particular interest. At high energy, the dissociation of the virtual photon can be interpreted as the diffractive fragmentation of a quark-antiquark dipole state, in the so-called {\em dipole picture} of DIS (see~\cite{kovchegov.levin.2012} for a review). The latter can be described using an elegant formulation~\cite{kovchegov.levin.2000,kovchegov.2012} in the form of QCD nonlinear evolution equations of cross sections. In this work, we employ the aforementioned dipole picture of DIS and numerical solutions to nonlinear evolution equations in both fixed and running coupling cases to study the diffractive dissociation of the virtual photon in the scattering off a large nucleus, with the main attention being on the rapidity gap distribution. Our motivation is twofold. First, the existence of the rapidity gap is the distinguishing feature of diffractive events, and it is used to mark such events in practice. Second, the size of the rapidity gap is, as suggested by Refs.~\cite{mueller.munier.2018a,mueller.munier.2018b,LMM.2021a}, the imprint of the partonic structure of the virtual photon subject to high-energy evolution. Therefore, the rapidity gap distribution is an important observable to measure in future electron-ion machines, as it is relevant to the microscopic mechanism of diffraction.

The paper is outlined as follows. In the next section, we will briefly introduce the dipole formulation for diffractive dissociation in DIS. In \cref{sec:diff_crsec}, the numerical evaluation of the diffractive cross section and of the rapidity gap distribution will be presented. Finally, we will draw some conclusions in \cref{sec:conclusion}.

\section{Dipole formulation for diffractive dissociation}
\label{sec:formalism}

In an appropriate frame, the virtual photon ($\gamma^*$) in high-energy DIS converts into a quark-antiquark pair, which is called {\em onium} hereafter, long before the scattering with the nucleus ($A$). Consequently, the photon cross sections can be expressed as a factorization of the photon wave function and the dipole cross sections in the scattering off a nucleus. Assuming impact parameter independence, the ratio of the diffractive cross section to the total cross section for diffractive $\gamma^* A$ scattering events with a {\em minimal} rapidity gap $Y_0$ out of the total rapidity $Y$ reads
\begin{equation}	\label{eq:tot_crsection}
	\frac{\sigma_{\rm diff}^{\gamma^* A}(Q^2,Y,Y_0)}{\sigma_{\rm tot}^{\gamma^* A}(Q^2,Y)} = \frac{\int d^2\underline{r} \int_0^1 dz \sum_{k=L,T;f}|\psi_k^{f}(\underline{r},z,Q^2)|^2\left[ 2N(r,Y) - N_{in}(r,Y,Y_0)\right]}{\int d^2\underline{r} \int_0^1 dz \sum_{k=L,T;f}|\psi_k^{f}(\underline{r},z,Q^2)|^2\ 2N(\underline{r},Y)},
\end{equation}
where $Q^2$ is the virtuality of the photon, $\underline{r}$ is the size of the onium in the two-dimensional transverse plane, and $z$ is the longitudinal momentum fraction of the photon taken by the quark. The probability densities $|\psi_{L,T}^{f}(\underline{r},z,Q^2)|^2$ of the photon-to-onium splitting in the longitudinal (L) and transverse (T) polarizations are well-known and can be found, for e.g., in Ref.~\cite{kovchegov.levin.2012}.
%
%\begin{equation}	\label{eq:wave_function}
%\begin{split}
%	|\psi_L^{f}(\underline{r},z,Q^2)|^2 & = \frac{\alpha_{\rm em}N_c}{2\pi^2} 4Q^2 z^2(1-z)^2 e_f^2 K_0^2(ra_f), \\ 
%	|\psi_T^{f}(\underline{r},z,Q^2)|^2 & = \frac{\alpha_{\rm em}N_c}{2\pi^2} e_f^2\left\{a_f^2K_1^2(ra_f)\left[z^2+(1-z)^2\right] + m_f^2K_0^2(ra_f)\right\},
%\end{split}
%\end{equation}
%
%where $r = |\underline{r}|$, $a_f^2 = Q^2z(1-z) + m_f^2$, and $m_f$ and $e_f$ are, correspondingly, the mass and the electric charge of a quark flavor $f$. 

%For the nuclear scattering of an onium of size $\underline{r}$, the total cross section $\sigma^{\rm onium}_{\rm tot}(\underline{r},Y)$ is related to the forward elastic scattering amplitude $N(r,Y)$ by
%
%\begin{equation}	\label{eq:sigma_1}
%	\sigma^{\rm onium}_{\rm tot}(\underline{r},Y) = \sigma_0 ,
%\end{equation}
%
%while the diffractive cross section can be computed using $N(r,Y)$ and the %amplitude $N_{\rm in}(r,Y,Y_0)$ encoding all inelastic contributions as
%
%\begin{equation}	\label{eq:sigma_2}
%	\sigma^{\rm onium}_{\rm diff}(\underline{r},Y,Y_0) = \sigma_0.
%\end{equation}
%
In \cref{eq:tot_crsection}, the function $N(r,Y)$ is the forward elastic scattering amplitude, and $N_{in}(r,Y,Y_0)$ is the cross section encoding all elastic contributions of the nuclear scattering of a dipole of transverse size $r$. They both obey the following leading-order (LO) Balitsky-Kovchegov (BK) equation~\cite{kovchegov.levin.2000,balitsky.1996,kovchegov.1999}:
\begin{equation}	\label{eq:bk_lo}
	\partial_Y \mathfrak{N}_r =  \int dp^{LO}(\underline{r},\underline{r'}) \left[\mathfrak{N}_{r'}+\mathfrak{N}_{|\underline{r}-\underline{r'}|}-\mathfrak{N}_{r}-\mathfrak{N}_{r'}\mathfrak{N}_{|\underline{r}-\underline{r'}|}\right], 
\end{equation}
with $\mathfrak{N}_r$ representing $N(r,Y)$ or $N_{in}(r,Y,Y_0)$. The LO integral kernel $dp^{LO}(\underline{r},\underline{r'})$ is given by
\begin{equation}	\label{eq:lo_kernel}
	dp^{LO}(\underline{r},\underline{r'}) = \frac{\bar{\alpha}}{2\pi}  \frac{r^2}{{r'}^2|\underline{r}-\underline{r'}|^2} d^2\underline{r'},
\end{equation}
where the ``reduced" strong coupling $\bar{\alpha} \equiv \frac{\alpha_sN_c}{\pi}$ is kept fixed (hereafter denoted by {\em fc}). The running-coupling extension of the BK equation (\ref{eq:bk_lo}) is also available in the literature~\cite{albacete.kovchegov.2007,balitsky.2007,kovchegov.weigert.2007,kovchegov.2012}, with the LO kernel (\ref{eq:lo_kernel}) being replaced by an appropriate kernel, depending on the prescription. In the current study, we employ the two following kernels:
\begin{enumerate} [label=(\roman*)]
	\item the ``parent dipole" kernel (denoted by {\em rc:pd})~\cite{albacete.kovchegov.2007}:
	\begin{equation}
	dp^{rc:pd} (\underline{r},\underline{r'}) = \frac{\bar{\alpha}(r^2)}{2\pi}\frac{r^2}{{r'}^2|\underline{r}-\underline{r'}|^2} d^2\underline{r'},
	\label{eq:pd_kernel}
	\end{equation}
	\item the Balitsky kernel (denoted by {\em rc:bal})~\cite{balitsky.2007}: 
	\begin{multline}
	dp^{rc:Bal} (\underline{r},\underline{r'}) = \frac{\bar{\alpha}(r^2)}{2\pi}\left[ \frac{r^2}{{r'}^2|\underline{r}-\underline{r'}|^2} + \frac{1}{{r'}^2} \left( \frac{\bar{\alpha}({r'}^2)}{\bar{\alpha}(|\underline{r}-\underline{r'}|^2)} - 1\right) \right.\\ 
	\left. + \frac{1}{|\underline{r}-\underline{r'}|^2} \left( \frac{\bar{\alpha}(|\underline{r}-\underline{r'}|^2)}{\bar{\alpha}({r'}^2)} - 1\right) \right] d^2\underline{r'}.
	\label{eq:Balitsky_kernel}
	\end{multline}
\end{enumerate}
In those prescriptions, the strong coupling $\bar{\alpha}(r^2)$ is set to run with size $r$ as in Ref.~\cite{albacete.kovchegov.2007}.

The BK equation can be solved numerically once the initial condition is set. The amplitude $N$ is usually initialized at $Y=0$ as the McLerran-Venugopalan (MV) amplitude~\cite{mclerran.venugopalan.1994a,mclerran.venugopalan.1994b},
\begin{equation}	\label{eq:mvi}
	N_{MV}(r,Y=0) = 1-\exp\left[-\frac{r^2Q_A^2}{4}\ln\left(e+\frac{1}{r^2\Lambda_{QCD}^2}\right)\right],
\end{equation}
with the nuclear saturation scale at zero rapidity $Q_A$ assumed to scale as $Q_A^2=0.26A^{1/3}\Lambda_{QCD}^2$. Meanwhile, the initial condition for $N_{in}$ is defined at $Y=Y_0$ as
\begin{equation}	\label{eq:init_Nin}
	N_{in}(r,Y=Y_0,Y_0) = 2N(r,Y_0)-N^2(r,Y_0).
\end{equation}
The detailed numerical recipe can be found in Ref.~\cite{Le.2021}.

We are interested in the rapidity gap distribution, which is defined as
\begin{equation}	\label{eq:gap_distrib_photon}
	\Pi^{\gamma^*A}(Q^2,Y;Y_{\rm gap}) \equiv \left|\frac{1}{\sigma_{\rm tot}^{\gamma^*A}}\frac{\partial \sigma_{\rm diff}^{\gamma^*A}}{\partial Y_0}\Bigr|_{Y_0=Y_{\rm gap}}\right|
\end{equation}
in the case of virtual photon-nucleus scattering, and 
\begin{equation}	\label{eq:gap_distrib_onium}
	\Pi^{\rm onium}(r,Y;Y_{\rm gap}) \equiv \left|\frac{1}{2N}\frac{\partial N_{in}}{\partial Y_0}\Bigr|_{Y_0=Y_{\rm gap}}\right|
\end{equation}
for the nuclear scattering of an onium.

An approach for the latter has been proposed recently~\cite{mueller.munier.2018a,mueller.munier.2018b,LMM.2021a}, in the case in which the onium is much smaller than the inverse nuclear saturation scale at the total rapidity $1/Q_s(Y)$. The main conjecture is that diffraction is due to a rare fluctuation in the Fock state of the onium which creates a parton with an unusual small transverse momentum. Within this picture, the asymptotic gap distribution reads~\cite{mueller.munier.2018a,mueller.munier.2018b,LMM.2021a}
\begin{equation}	\label{eq:asymp_distrib}
	\Pi^{\rm onium}_{\infty}(r,Y;Y_{\rm gap}) = c_D \left[\frac{Y}{Y_{\rm gap}(Y-Y_{\rm gap})}\right]^{3/2},
\end{equation}
valid in the so-called ``scaling region'' $1\ll \ln \frac{1}{r^2Q_s^2(Y)} \ll \sqrt{Y}$ and with the constant $c_D$ completely determined~\cite{LMM.2021a}.

\section{Numerical analysis of diffraction}
\label{sec:diff_crsec}

\begin{figure}[h!]
	\centering
	\includegraphics[width=0.98\textwidth]{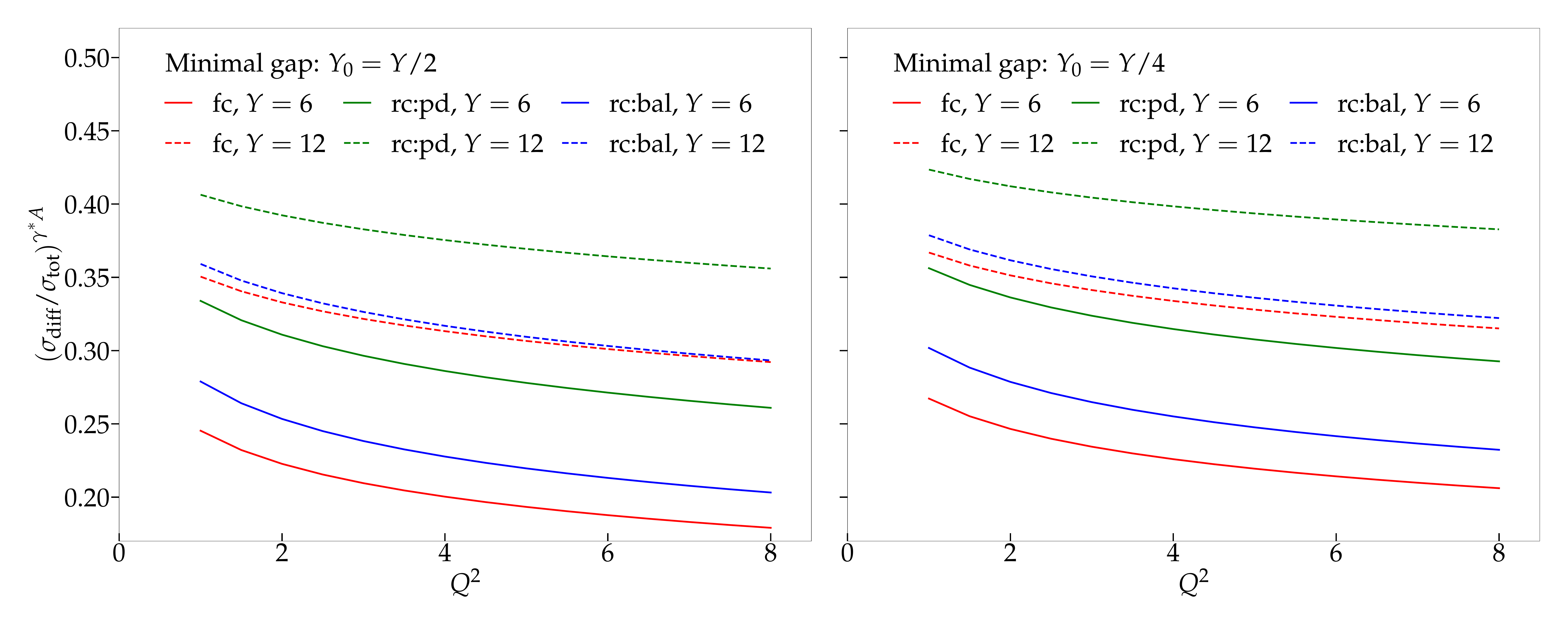}
	\caption{The diffractive fraction for two different values of the minimal rapidity gap $Y_0$ and two values of the total rapidity $Y$ as a function of the virtuality $Q^2$. $A=208$ is set for the nuclear mass number.}
	\label{fig:diffractive_ratio}
\end{figure}

The ratio of the diffractive cross section to the total cross section, or the diffractive fraction for short, is plotted in~\cref{fig:diffractive_ratio} (see Ref.~\cite{Le.2021} for a complete definition of the setting and parameters). While the fraction of diffractive events, about $20\%-25\%$ in the fixed coupling scheme and for a low rapidity ($Y=6$), is rather close to the values predicted by other studies (for example, see Ref.~\cite{Bendova.etal.2020}), the results for other cases seem to overshoot the latter.

Different behaviors of the diffractive fraction can be drawn from~\cref{fig:diffractive_ratio}. First, the ratio decreases with $Q^2$, while it grows when boosting to a higher total rapidity $Y$. These two features reflect the dependence of the diffractive fraction on the ratio between the virtuality $Q$ and the saturation scale $Q_s(Y)$. In particular, the smaller the scale ratio $Q/Q_s(Y)$ is, the more deeply the process probes in the saturation region. Consequently, the diffractive fraction gets closer to its black-disk limit value ($0.5$). Furthermore, the running coupling correction produces larger values for the fraction. This can be attributed to the effect of the running coupling to suppress the emission of gluons with large transverse momentum (equivalently, the branching to color dipoles of small transverse sizes in the limit of large $N_c$) in the evolution of the onium state of the photon. The process is therefore more elastic, and closer to the black-disk limit.

\begin{figure}[h!]
	\centering
	\includegraphics[width=0.97\textwidth]{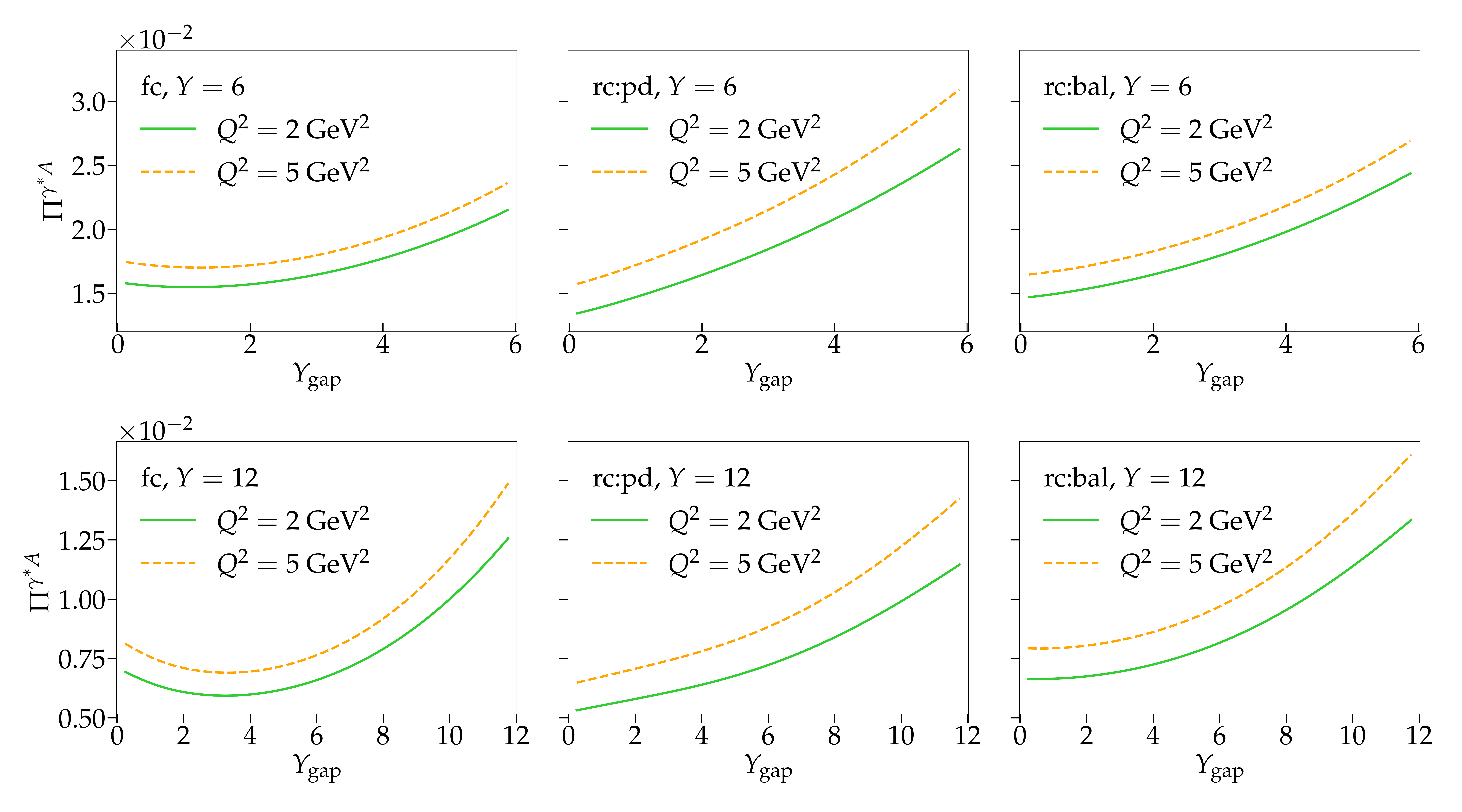}
	\caption{Rapidity gap distribution for different values of the total rapidity $Y$ and of the virtuality $Q^2$. $A=208$ is set for the nuclear mass number.}
	\label{fig:gap_dist}
\end{figure}
Figure \ref{fig:gap_dist} shows the rapidity gap distributions for different scenarios. There is a difference in the general shape of the distribution between the fixed coupling and running coupling cases. In particular, there is a local minium located at a gap value $0<Y_{\rm gap}<Y$, while the distribution is enhanced for the gap values close to $0$ or $Y$. That is not the case when the running coupling effect is taken into account. For the latter, the distributions appear to grow monotonically when the size of the rapidity gap increases. Furthermore, one should notice a notable feature that the shape of the distribution when the strong coupling is fixed is quite similar to the shape predicted by Refs.~\cite{mueller.munier.2018a,mueller.munier.2018b,LMM.2021a} (see \cref{eq:asymp_distrib}) in the case of onium-nucleus scattering.

\section{Conclusion}
\label{sec:conclusion}

We have presented numerical predictions for the diffractive fraction and the rapidity gap distribution for kinematics accessible at future elecron-ion colliders, based on the solutions to QCD small-$x$ evolution equations in both fixed coupling and running coupling scenarios. The diffractive events are shown to take a significant fraction in the scattering of a virtual photon off a large nucleus. The shape of the rapidity gap distribution is significantly modified by the inclusion of running coupling corrections. Interestingly, the distribution in the fixed coupling case has a shape which is qualitatively similar to the shape deduced from a recently proposed partonic model. This suggests that the distribution of the gap size could be a relevant observable for unveiling the microscopic mechanism of  diffractive dissociation.

The current work takes into account the running coupling effect, which is the only-known next-to-leading correction thus far for the diffractive dissociation. In addition, the knowledge on the rapidity gap distribution in the onium-nucleus scattering is currently limited to the regime of asymptotic large rapidity and small onium size in the scaling region. Therefore, to produce better phenomenological predictions for electron-nucleus collision, further theoretical developments are essential.

\section*{Acknowledgements}
We would like to thank Stéphane Munier for his valuable comments and for reading the manuscript. 

% TODO: include funding information
\paragraph{Funding information}
This work is supported in part by the Agence Nationale de la Recherche under the project ANR-16-CE31-0019.

% TODO:
% Provide your bibliography here. You have two options:

% FIRST OPTION - write your entries here directly, following the example below, including Author(s), Title, Journal Ref. with year in parentheses at the end, followed by the DOI number.
%\begin{thebibliography}{99}
%\bibitem{1931_Bethe_ZP_71} H. A. Bethe, {\it Zur Theorie der Metalle. i. Eigenwerte und Eigenfunktionen der linearen Atomkette}, Zeit. f{\"u}r Phys. {\bf 71}, 205 (1931), \doi{10.1007\%2FBF01341708}.
%\bibitem{arXiv:1108.2700} P. Ginsparg, {\it It was twenty years ago today... }, \url{http://arxiv.org/abs/1108.2700}.
%\end{thebibliography}

% \bibliographystyle{SciPost_bibstyle} % Include this style file here only if you are not using our template
%\twocolumn
\bibliography{bib.bib}

%\nolinenumbers

\end{document}